\documentclass[reprint, superscriptaddress, secnumarabic, amssymb, nobibnotes, aps, prl]{revtex4-1}

\usepackage{graphicx}
\usepackage{epstopdf}
\usepackage[T1]{fontenc}
\usepackage[latin9]{inputenc}
\usepackage{amsbsy}
\usepackage{gensymb}
\usepackage[T1]{fontenc}
\usepackage[latin9]{inputenc}
\usepackage{amsmath}
\usepackage{amssymb}
\usepackage{bbm}
\usepackage{braket}
\usepackage{xcolor}
\allowdisplaybreaks
\usepackage{graphicx}
\usepackage[colorlinks=true]{hyperref}  
\hypersetup{
    bookmarks=true,         
    unicode=false,          
    pdftoolbar=true,        
    pdfmenubar=true,        
    pdffitwindow=false,     
    pdfstartview={FitH},    
    pdftitle={HEA_2${nd}$},    
    pdfauthor={},     
    pdfsubject={},   
    pdfcreator={},   
    pdfproducer={}, 
    pdfkeywords={} {} {}, 
    pdfnewwindow=true,      
    colorlinks=true,       
    linkcolor=blue, 
    citecolor=blue,        
    filecolor=magenta,      
    urlcolor=blue           
} 
\usepackage[normalem]{ulem}

\begin{document}

\title{\textrm{Boron based new high entropy alloy superconductor Mo$_{0.11}$W$_{0.11}$V$_{0.11}$Re$_{0.34}$B$_{0.33}$ }}
\author{Kapil Motla}
\affiliation{Department of Physics, Indian Institute of Science Education and Research Bhopal, Bhopal, 462066, India}
\author{Arushi}
\affiliation{Department of Physics, Indian Institute of Science Education and Research Bhopal, Bhopal, 462066, India}
\author{Pavan K. Meena}
\affiliation{Department of Physics, Indian Institute of Science Education and Research Bhopal, Bhopal, 462066, India}
\author{R. P. Singh}
\email[]{rpsingh@iiserb.ac.in}
\affiliation{Department of Physics, Indian Institute of Science Education and Research Bhopal, Bhopal, 462066, India}

\begin{abstract}
\begin{flushleft}
\end{flushleft}
Superconducting high entropy alloys (HEAs) are new members of disordered superconductors. We report the synthesis and investigation of a new superconducting high entropy alloy Mo$_{0.11}$W$_{0.11}$V$_{0.11}$Re$_{0.34}$B$_{0.33}$ (MWVRB). It crystallized in the tetragonal CuAl$_2$ crystal structure with space group (I4/$mcm$). Comprehensive transport, magnetization and heat capacity measurements confirmed bulk type-II superconductivity having transition temperature T$_{C}$ = 4.0 K. The low temperature electronic specific heat suggests a fully gapped superconducting state in weak coupling limit.
\end{abstract}

\maketitle

\section{Introduction}
High entropy alloys are getting enormous interest in material science due to their unique composition and outstanding properties over conventional alloys \cite{Cantor,E P George Nature,Most required}. The alloy contains at least five constituent elements with 5 to 35 atomic $\%$ is called high entropy alloy \cite{nano, DB, Y Zou, Yeh}, and their inherent feature "high entropy" help to crystallize the random solid solution in simple crystallographic lattice such as in bcc \cite{prl}, fcc \cite{fcc}, $\alpha$-Mn \cite{cava,muon hea} and hcp \cite{Hexa} crystal structure \cite{prx,nano}. Superconductivity was also observed in some high entropy alloys despite their high disordered nature, where the possibility of regular phonon modes are unlikely \cite{phonon broadening}. The superconducting phenomenon combined with mechanical properties such as high thermal stability, high strength, and excellent corrosion resistance exhibited in HEAs makes them a promising system for fundamental study as well as for application purposes \cite{Tasan,thermal stability,strength,Corrosion}. These high entropy alloy superconductors were reported to show some unconventional properties like retention of superconductivity at extremely high pressure, anomalous broadening in specific heat jump, lattice heat parameter in the elemental range \cite{pressure hea, anomalous broadening} and superconducting parameters similar to phonon mediated superconductors \cite{muon hea}. Understanding the occurrence of superconductivity and exotic properties in these highly disordered and multicomponent alloys remains challenging due to the lack of detailed electronic structure and lattice vibration, which is vital to understanding the superconducting pairing mechanism. The chemical complexity and disorder in the HEAs can provide a versatile platform to investigate the relationship between the property of an ordered to disordered state \cite{Complexity}. Currently, a lot of effort is going on to discover new superconducting high entropy alloys in different crystal structures and tune their superconducting transition temperature using the combination of 3d, 4d, and 5d elements. Same time very few microscopic studies and the unavailability of different families of high entropy alloy superconductors make it difficult to understand the superconducting pairing mechanism \cite{muon hea}. It is clearly important to search for new HEA superconductors to understand the superconducting ground state properties of these emerging new families of disordered superconductor.\\
In this paper, we report the synthesis and a detailed investigation of the superconductivity in rhenium and boron-rich new HEA superconductor using magnetization, resistivity and heat capacity. It crystallized in a tetragonal CuAl$_2$ crystal structure with space group (I4/${mcm}$) and having a superconducting transition temperature 4.0 $K$. Uemura plot constructed using electronic property calculations places MWVRB in the band of conventional superconductors.
\section{Experimental Details}
A polycrystalline sample of MWVRB was synthesized using arc melting stoichiometric quantities of Mo(99.95$\%$), W(99.95$\%$), V(99.7$\%$), Re(99.99$\%$), and B(99.5$\%$) in a single arc furnace under high purity argon gas. A Ti getter is used to remove any residual oxygen present in the chamber. The ingot was flipped and remelted several times to ensure sample homogeneity. The resulting sample was shiny, and weight loss was negligible. The phase purity of the sample was confirmed by powder X-ray diffraction on a PANalytical X-Pert Pro diffractometer using Cu-K$_{\alpha}$, ($\lambda$ = 1.5405 \text{\AA}) radiation at ambient temperature. Macroscopic superconductivity was verified by DC magnetization using a vibrating sample magnetometer (VSM) option of the Quantum Design MPMS 3 (Magnetic Property Measurement System). Transport and heat capacity measurements were carried out using the Quantum Design physical property measurement system (PPMS).\\
\begin{figure} 
\centering
\includegraphics[width=1.0\columnwidth, origin=b]{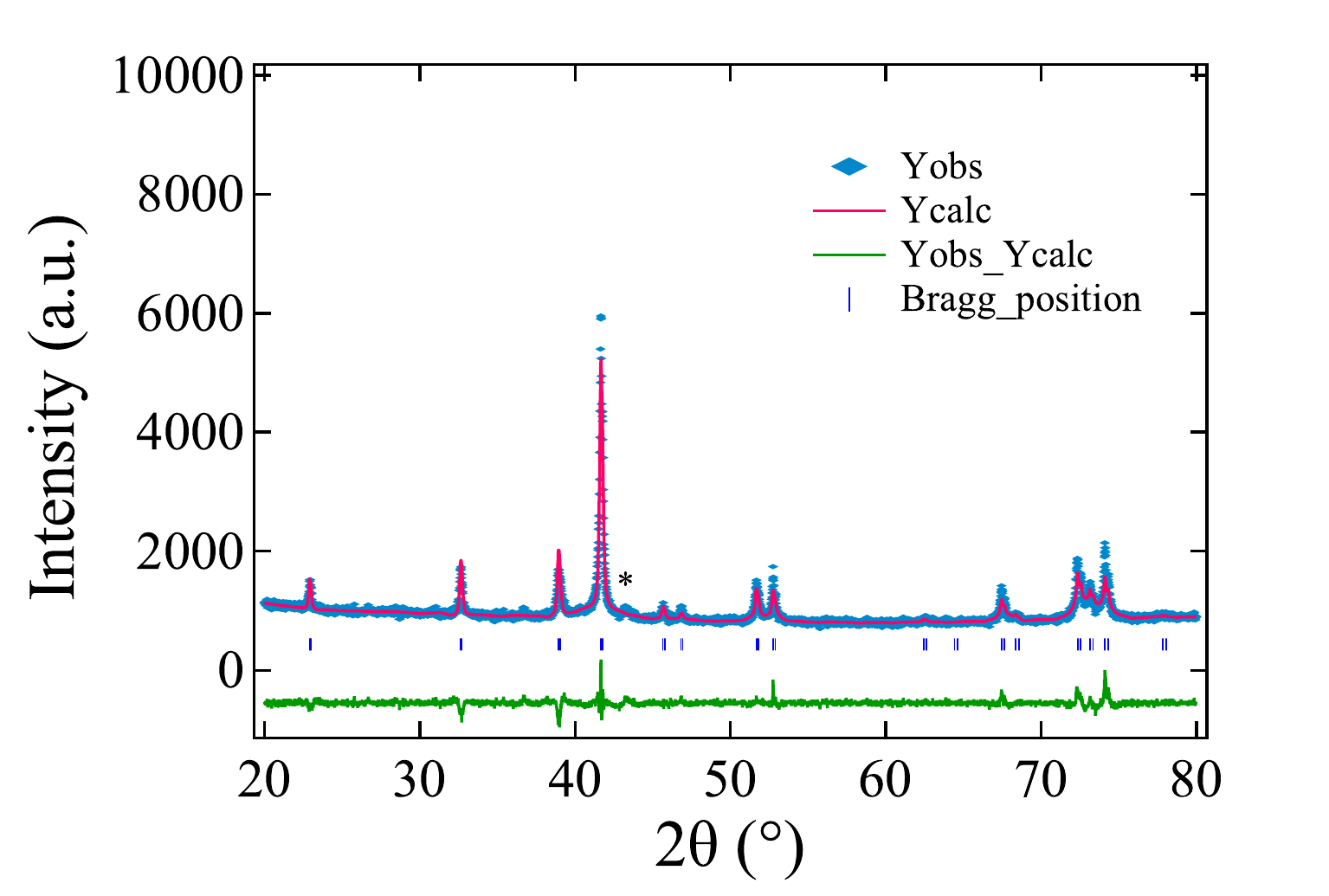}
\caption{Powder X-ray diffraction pattern collected at room temperature using CuK$_{\alpha}$ radiation. The observed pattern is fitted with a tetragonal (I4/$mcm$) structure (solid red line).}
\centering
\label{Fig1}
\end{figure}
\section{Results and Discussion}
\subsection{Structural characterization}
The room temperature X-ray diffraction pattern of the as-cast MWVRB sample is shown in Fig. \ref{Fig1}. The Le-bail fitting was used to determine the crystal structure and lattice parameters by employing Fullprof software. It confirmed that the MWRVB formed in a tetragonal CuAl$_2$ type crystal structure (space group (I4/$mcm$)) with lattice parameters a = b = 5.4848 {\AA} and c = 4.6240 {\AA}. The broadness of the peak can be attributed to the degree of disorder in this HEA. Despite large atomic difference Mo$_{0.11}$W$_{0.11}$V$_{0.11}$Re$_{0.34}$B$_{0.33}$ (MWVRB) stabilized in a single phase. \\

\subsection {Electrical resistivity}
\begin{figure} 
\includegraphics[width=1.0\columnwidth, origin=b]{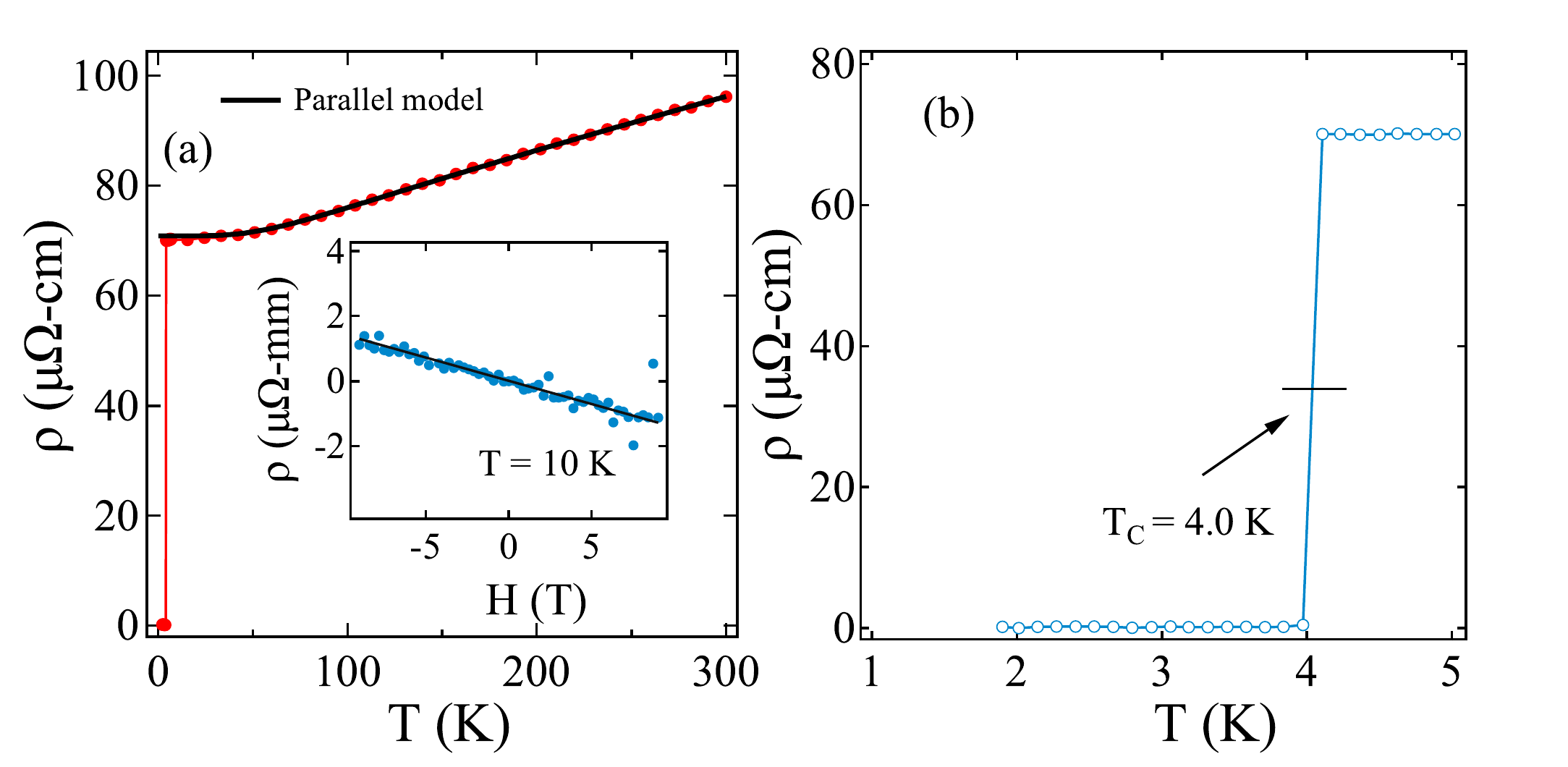}
\caption{(a) The temperature dependent resistivity of MWVRB sample in temperature range 1.9 $K$ $\leq$ $T$ $\leq$ 300 $K$ is fitted using parallel resistor model by solid black line, and the inset shows the Hall resistivity at 10 $K$ (> T$_{C}$). (b) The sharp superconducting transition is shown with T$_{C}$ = 4.0 $K$.}
\label{Fig2}
\end{figure}

Fig. \ref{Fig2}(a) shows the temperature-dependent resistivity $\rho$(T) in the temperature range of 1.9 K to 300 K in zero applied field, and the sharp transition is observed at T$_{C}$ = 4.0 $K$ [see Fig. \ref{Fig2}(b)]. The observed resistivity at room temperature is 71 $\mu \ohm cm$ and the residual resistivity ratio RRR = $\rho_{300K}$/$\rho_{7K}$ = 1.37 and is comparable with some Re-based binary and HEA superconductors \cite{Re24Ti5RRR,Nb0.18Re0.82RRR, muon hea}. The low value of RRR may be due to the existence of the atomic-scale disorder and the very small size of the domains in HEAs \cite{prl,maneesha,s wave hea}. 
$\rho(T)$ above 50 $K$ is found to be linear with temperature and then saturates at high temperatures. Similar trend was also seen in Re$_{6}$Zr \cite{Re6Zrparallel} and BiPd \cite{BiPd} and it is proposed that mean free path in some materials become shorter, which leads to the deviation in resistivity from linearity to saturation \cite{resistivity saturate}. This type of behaviour of $\rho(T)$ was well described using parallel resistor model and expressed as
\begin{equation}
 \rho(T) = \left[\frac{1}{\rho_{s}} + \frac{1}{\rho_{i}(T)} \right]^{-1}
\label{para1}
\end{equation}
where $\rho_{s}$ is the saturation resistivity achieved at higher temperatures and $\rho_{i}(T)$ is ideal temperature dependent resistivity given as
\begin{equation}
 \rho_{i}(T) = \rho_{i,0} + \rho_{i,L}(T)
\label{para2}
\end{equation}\\
where the first term $\rho_{i,0}$ is temperature-independent residual resistivity.  The second term is temperature-dependent, which accounts the phonon-assisted electron scattering and is expressed by the generalized BG resistivity model as
\begin{equation}
 \rho_{i,L}(T) = C\left(\frac{T}{\Theta_{D}}\right)^{5}\int_{0}^{\Theta_{D}/T}\frac{x^{5}}{(e^{x}-1)(1-e^{-x})}dx
\label{para3}
\end{equation}\\
here C is a material dependent quantity and $\Theta_{D}$ is the Debye temperature, which is obtained from resistivity measurements. The observed resistivity data is well fitted with theoretical red solid line and is shown in Fig. \ref{Fig2}(a). The fitting yields $\rho_{0}$ = 71.1 $\mu\ohm cm$, $\rho_{0,s}$ = 112 $\mu\ohm cm$, and $\Theta_{D}$ = 335 K, which is in good agreement with the $\Theta_{D}$ obtained from specific heat measurement (discussed later).

\subsection{Magnetization}
\begin{figure} 
\centering
\includegraphics[width=1.0\columnwidth, origin=b]{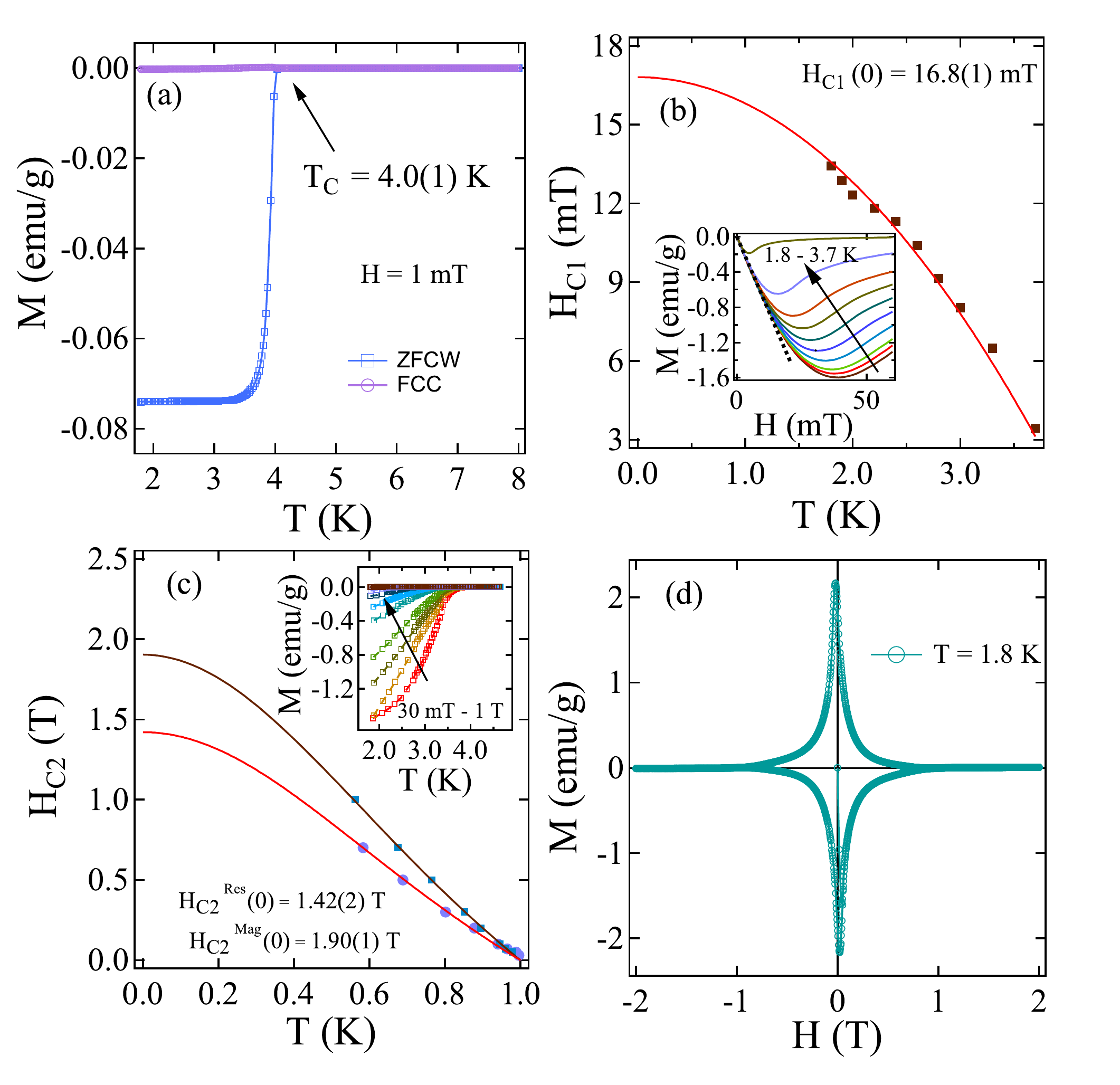}
\caption{(a) Temperature dependence of DC magnetic moment in both ZFCW and FCC mode under an applied field of 1 mT. (b) Temperature dependent lower critical field curve fitted with Eq. \ref{Hc1} and inset shows the field dependent magnetization within temperature range 1.8 K to 3.7 K. (c) Temperature dependence of upper critical field fitted with Eq. \ref{hc2} and inset represents magnetization $vs$ temperature within magnetic field range 30 mT to 1 T. (d) The field dependent magnetization M(H) at 1.8 $K$.}
\label{Fig3}
\end{figure}

To ensure bulk superconductivity in the MWVRB sample, DC magnetization measurements were performed in zero field-cooled warming (ZFCW) and field-cooled cooling (FCC) modes in 1 $mT$ magnetic field. The onset of diamagnetism is considered the transition temperature T$_{C}$ = 4.0(1) $K$, shown in Fig. \ref{Fig3}(a). In the superconducting state, FCC data indicate the strong pinning in MWVRB HEA, and above T$_{C}$, sample shows the paramagnetic nature. The temperature and magnetic field dependent magnetization of MWVRB was used to reveal several superconducting state parameters. The lower critical field was evaluated using the field-dependent magnetization (M-H) collected at different temperatures from base value to T$_{C}$. In the low field range, the field-dependent magnetization data varies linearly and deviates at a particular value of the magnetic field, the so-called lower critical field for that temperature. The evaluated critical field value increases with decreasing temperature, and the observed data is fitted well with the Ginzburg-Landau expression as 

\begin{equation}
H_{C1}(T)=H_{C1}(0)\left(1-\left(\frac{T}{T_{C}}\right)^{2}\right)
\label{Hc1}
\end{equation}
We extrapolate the theoretical curve up to 0$K$, which yields the H$_{C1}$(0)= 16.8(1) $mT$ as shown in Fig. \ref{Fig3}(b).\\
Another characteristic parameter, the upper critical field was calculated using the temperature dependent magnetization and resistivity measurements in varying applied magnetic fields [see Fig. \ref{Fig3}(c)]. The  $T_{C}^{onset}$ and $T_{C}^{mid}$ were the criteria of transition temperature for the upper critical field from magnetization and resistivity measurements, respectively. The onset of diamagnetism in magnetization curves shifted towards lower temperature with the application of increasing magnetic fields, as shown in the inset of Fig. \ref{Fig3}(c). The $H_{C2}(T)$ $vs$ $T$ data were explained very well by the Ginzburg-Landau expression 
\begin{equation}
H_{C2}(T) = H_{C2}(0)\left(\frac{1-t^{2}}{1+t^2}\right) 
\label{hc2}
\end{equation}
\\
where $t = T/T_{C}$ is the reduced temperature, and by extrapolate the theoretical curve up to 0$K$, we obtained $H_{C2}^{Mag, Res}$(0) = 1.90(1), 1.42(2) $T$ from magnetization and resistivity measurements, respectively. Furthermore, according to the expression, the upper critical field is directly related to the Gingburg-Landau coherence length as
\begin{equation}
H_{C2}(0) = \frac{\Phi_{0}}{2\pi\xi_{GL}^{2}}
\label{eqn7:up}
\end{equation}
where $\Phi_0$ is the magnetic quantum flux ($\Phi_0$ = 2.07$\times$10$^{-15}$T-m$^2$) and using the value of H$_{C2}$(0) = 1.90(1) $T$ (from magnetization), we obtained the coherence length $\xi_{GL}$(0) = 13.1(1) $nm$.
The lower critical field is related to the  magnetic penetration depth $\lambda_{GL}$(0) and is defined as 
\begin{equation}
H_{C1}(0) = \frac{\Phi_{0}}{4\pi\lambda_{GL}^2(0)}\left(\mathrm{ln}\frac{\lambda_{GL}(0)}{\xi_{GL}(0)}+0.12\right)   
\label{eqn6:ld}
\end{equation}
after substituting the values of $\xi_{GL}$(0) = 13.1(1) $nm$, and $H_{C1}$(0) = 16.8(1) mT, it yields the magnetic penetration depth $\lambda_{GL}$(0) = 160(11) $nm$, for MWVRB sample.\\

Ginzburg-Landau provide a parameter $\kappa_{GL}$, which can differentiate the types of superconductivity, and expressed in terms of penetration depth $\lambda_{GL}$(0) and coherence length $\xi_{GL}$(0) as $\kappa_{GL}$=$\frac{\lambda_{GL}(0)}{\xi_{GL}(0)}$, using $\xi_{GL}$(0) = 13.1(1) $nm$ and $\lambda_{GL}$(0) = 160(11) $nm$ for MWVRB, we have calculated the $\kappa_{GL}$ = 12.2(1) >> ${\frac{1}{\sqrt{2}}}$, which indicate that the MWVRB HEA is a type II superconductor. 
The impact of the applied magnetic field on Cooper pair breaking occurs via two types of mechanism; either the orbital field effect or the Pauli limiting magnetic field effect, which destroys the superconductivity of any material. Using the Werthamar-Helfand-Hohenberg (WHH) model without considering spin-orbit interaction \cite{whh}, the orbital limiting field in dirty limit BCS superconductor is expressed as 
\begin{equation}
H_{C2}^{orbital}(0) = -\alpha   T_{C}\left.\frac{dH_{C2}(T)}{dT}\right|_{T=T_{C}}
\label{eqn8:whh}
\end{equation}
 where $\alpha$ can take 0.693 value in case of dirty limit superconductivity [discussed later]. Near T$_{C}$, the variation of temperature dependent upper critical field ${\frac{dH_{c2}}{dT}}$ is estimated to be  -1.2(1) $T$, which yields the orbital limiting field $H_{C2}^{orb}$ = 3.3(2) $T$.
Another effect which suppresses the superconductivity is  Pauli-clogston limiting field $H_{p}$ and within the BCS theory it can be expressed as H$_{c2}^{P}$ = 1.84 $T_{C}$ \cite{pauli}. 
The calculated value of $H_{c2}^{P}$ = 7.3(2) $T$. 
The estimation of the relative strength of orbital and Pauli limiting field value of the upper critical field is done by determining the Maki's parameter $\alpha_{M}$ as 
$\alpha_{M} = \sqrt{2}H_{C2}^{orb}(0)/H_{C2}^{p}(0)$. 
The value of $\alpha_{M}$ is 0.7 for MWVRB sample, which indicates the significant influence of orbital limiting field over the Pauli paramagnetic field in destroying the superconductivity.\\
The magnetization hysteresis loop ($\pm$ 2T) was taken at 1.8 $K$ and shown in Fig. \ref{Fig3}(d). At a value of magnetic field ($\pm$1T), a closed loop was observed and is called H$_{irr}$. The value of H$_{irr}$ is below the upper critical field, which is the indication of deppining of vortex flux line in MWVRB HEA. 
\begin{figure} 
\centering
\includegraphics[width=1.0\columnwidth, origin=b]{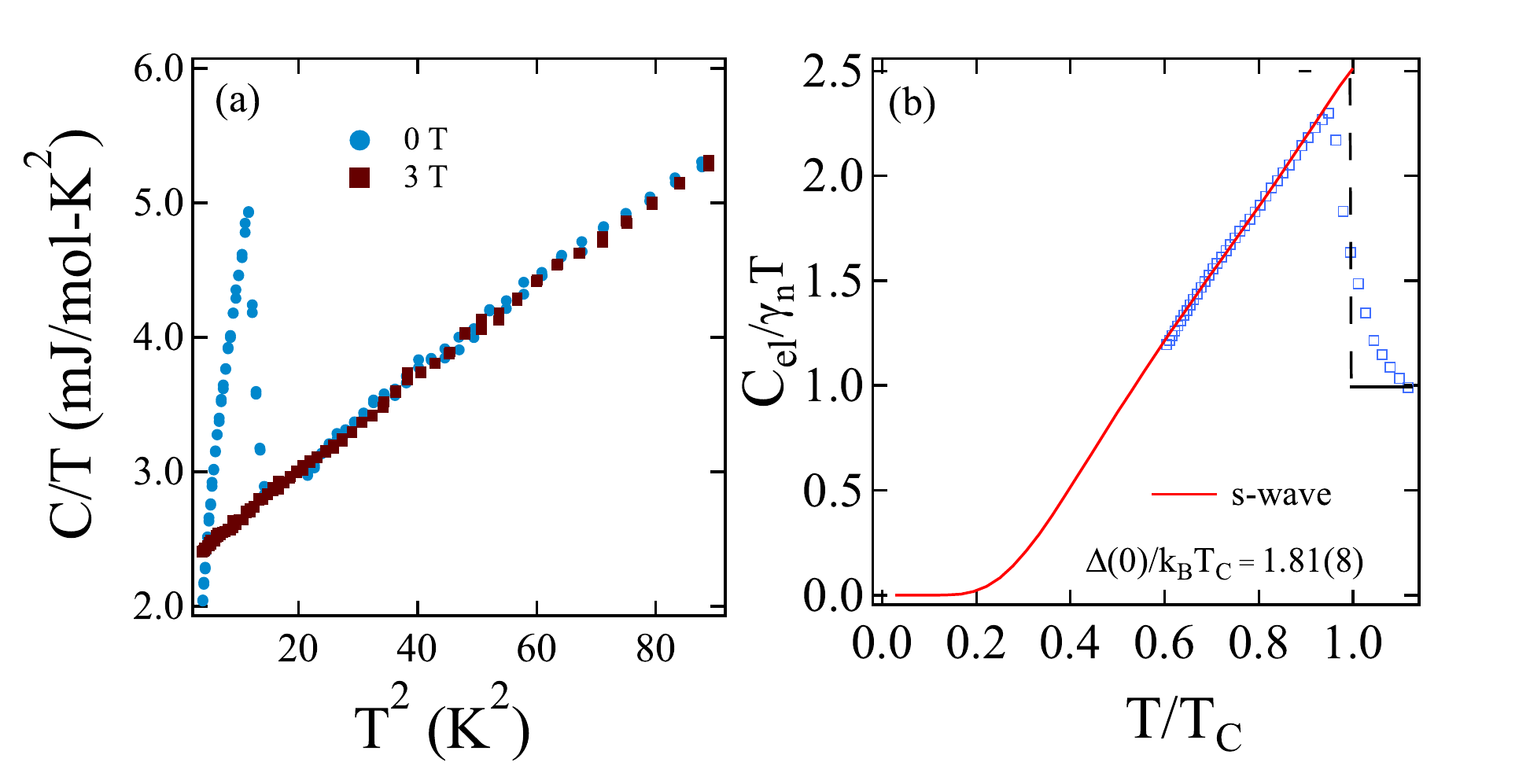}
\caption{(a) C/T $vs$ T$^2$ curve at 0 $T$ and 3 $T$ are shown by solid circle and square symbols, respectively. (b) The low temperature specific heat data in the measured temperature range is fitted using the BCS s-wave model.}
\label{Fig4}
\end{figure}
\subsection{Specific heat}
We have performed zero-field specific heat measurement to confirm bulk superconductivity and loss of entropy at T$_{C}$ in the new superconducting MWVRB HEA. The sharp jump in the specific heat $C/T$ $vs$ T$^{2}$ data is observed at the bulk superconducting transition temperature T$_{C}$ $\sim$ 3.6 $K$ as shown in Fig. \ref{Fig4}(a), which coincides with the superconducting transition temperature observed from magnetization and resistivity measurements. Above T$_{C}$, the ${C/T}$ ${vs}$ $T^{2}$ data is fitted well with the conventional Debye model as 
\begin{equation}
\frac{C}{T} = \gamma_{n} + \beta_{3}T^{2}
\label{eqn10:C}
\end{equation}
where $\gamma_{n}T$ and $\beta_{3}T^{3}$ are the electronic and the phononic heat capacity contributions, respectively. The normal state specific heat fitting provides the electronic heat capacity coefficient $\gamma_n$ = 2.41(4) $mJ$-$mol^{-1} K^{-2}$ (Sommerfeld parameter) and lattice heat capacity coefficient $\beta_3$ = 0.027(1) $mJ$-$mol^{-1}K^{-4}$ for MWVRB HEA. The $\beta_{3}$ coefficient is related to Debye temperature as 
\begin{equation}
\Theta_{D} = \left(\frac{12\pi^{4}RN}{5\beta_{3}}\right)^{\frac{1}{3}}
\label{eqn11:theta}
\end{equation}
where $R$ is the universal gas constant, and $N$ is the number of atoms per unit cell, which is 1. After substituting the related parameters, we calculated the Debye temperature $\Theta_{D}$ = 414(7) $K$. The Debye temperature is higher than conventional alloys and in the range of other high entropy alloy superconductors \cite{cava, muon hea}.
The electronic heat capacity constant is proportional to the density of state for the non-interacting system by the expression
\begin{equation}
\gamma_{n} = \left(\frac{\pi^{2}k_{B}^{2}}{3}\right)D_{C}\left(E_{F}\right)
\label{eqn12:gamma}
\end{equation}
here $k_{B}$ = 1.38$\times$ $10^{-23} JK^{-1}$ is the Boltzmann constant and the estimated value of density of state at the Fermi level D$_{C}(E_{F}$) is 1.02(2) $\frac{states}{eV f.u}$. 
The  electron-phonon interaction parameter $\lambda_{e-ph}$, which gives information regarding the  strength between electron and phonon coupling. The $\lambda_{e-ph}$ is related to $\Theta_{D}$ and $T_{C}$, and given by the McMillan model as \cite{Mcmillan}
\begin{equation}
\lambda_{e-ph} = \frac{1.04+\mu^{*}\mathrm{ln}(\theta_{D}/1.45T_{C})}{(1-0.62\mu^{*})\mathrm{ln}(\theta_{D}/1.45T_{C})-1.04 }
\label{eqn13:ld}
\end{equation}
where $\mu^{*}$ is the screened Coulomb potential, and the typical value is 0.13 for the intermetallic compound. Using $\Theta_{D}$ = 414(7) K, $T_{C}$ = 3.6 $K$ (from specific heat measurement), the calculated electron-phonon coupling parameter is $\lambda_{e-ph}$ = 0.54(7) that indicates weakly coupled  superconductivity in MWVRB system.\\
In order to estimate the superconducting gap structure and magnitude, the electronic specific heat $C_{el}$ can be estimated by excluding the phononic contribution from the total specific heat $C(T)$ using the expression  C$_{el}$ = $C(T)$ - $\beta_{3}T^3$.
The electronic specific heat data below T$_{C}$ can be well fitted with isotropic single-gap BCS model for normalized entropy S as
\begin{equation}
\frac{S}{\gamma_{n}T_{C}} = -\frac{6}{\pi^2}\left(\frac{\Delta(0)}{k_{B}T_{C}}\right)\int_{0}^{\infty}[ \textit{f}\ln(f)+(1-f)\ln(1-f)]dy \\
\label{eqn14:s}
\end{equation}
\\
where $\textit{f}$($\xi$) = [exp($\textit{E}$($\xi$)/$k_{B}T$)+1]$^{-1}$ is the Fermi function, $\textit{E}$($\xi$) = $\sqrt{\xi^{2}+\Delta^{2}(t)}$, where $E(\xi $) is the energy of the normal electrons relative to the Fermi energy, $\textit{y}$ = $\xi/\Delta(0)$, $\mathit{t = T/T_{C}}$ and $\Delta(t)$ = $tanh$[1.82(1.018(($\mathit{1/t}$)-1))$^{0.51}$] is the approximation of temperature dependent energy gap by  BCS theory. The normalized electronic specific heat in superconducting state is related with the normalized entropy as
\\
\begin{equation}
\frac{C_{el}}{\gamma_{n}T_{C}} = t\frac{d(S/\gamma_{n}T_{C})}{dt} \\
\label{eqn15:Cel}
\end{equation}
\\
The observed electronic heat capacity data below $T_{C}$ is fitted with Eq. \ref{eqn15:Cel} and is shown in Fig. \ref{Fig4}(b). The fitting in the measured temperature region provide $\Delta(0)/k_{B}T_{C}$ = 1.81(8). The value of gap with $\lambda_{e-ph}$ indicating the possibility of weakly coupled isotropic superconducting gap in MWVRB HEA. \\
\begin{figure}
\includegraphics[width=1.0\columnwidth, origin=b]{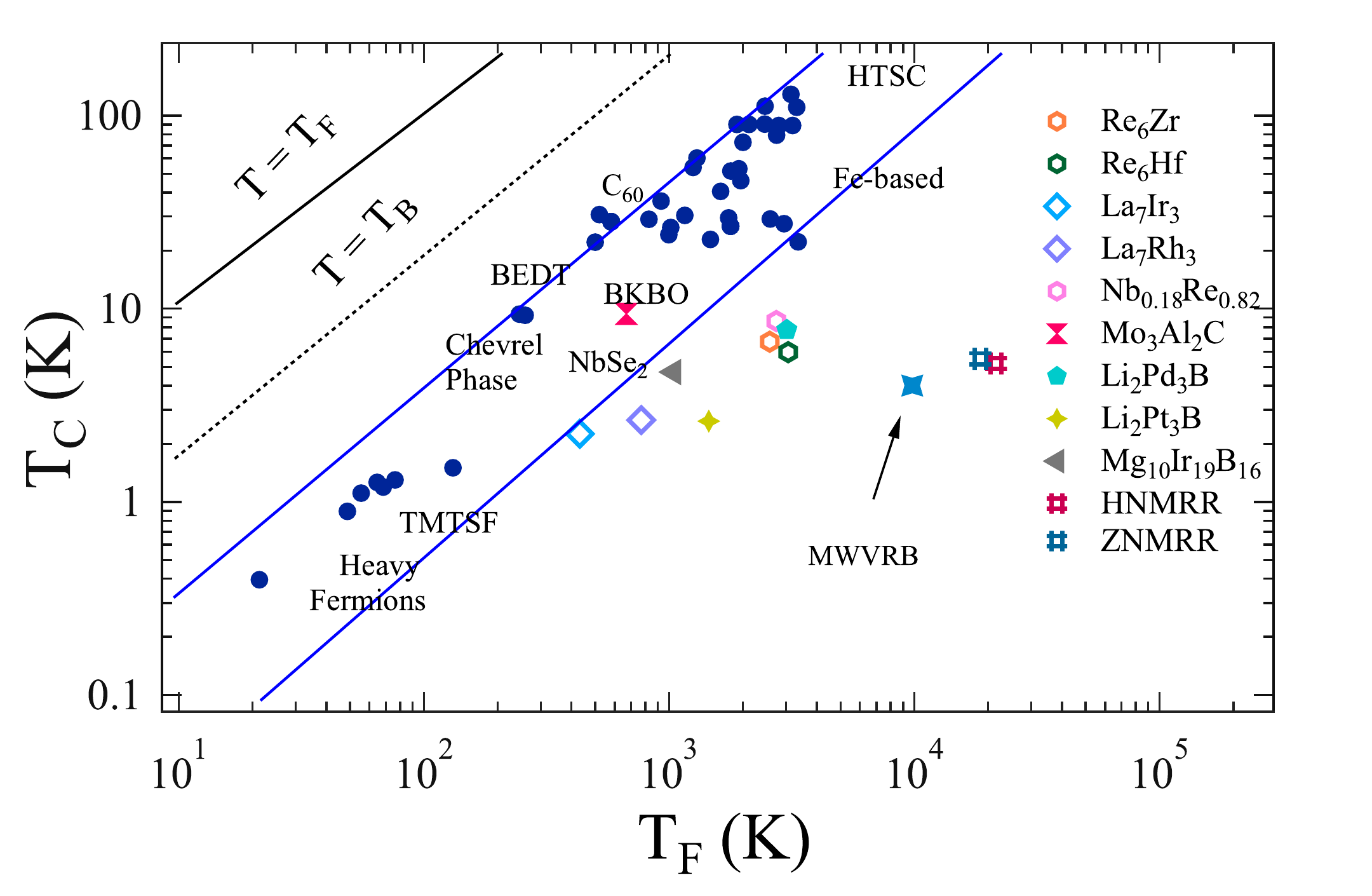}
\caption{A plot between the superconducting transition temperature T$_{C}$ and the Fermi temperature T$_{F}$ is shown for different superconducting families \cite{U3,elec-prop,muon hea}. The solid blue marker represents the position of MWVRB superconductor which is away from the unconventional band (region between two blue lines).}
\label{Fig5}
\end{figure}
\subsection{Electronic property and Uemura plot}
To obtain the electronic properties such as effective mass, mean free path, and Fermi velocity, we have used some calculated parameters such as charge density $n$ (from hall measurement), Sommerfeld coefficient $\gamma_{n}$ (by specific heat measurement), and the residual resistivity $\rho_{0}$ (from resistivity measurement). The carrier density $n$ and effective mass of quasi-particles $m^{*}$ are related to Sommerfeld coefficient $\gamma_{n}$ as \cite{Thinkam}
\begin{equation}
\gamma_{n} = \left(\frac{\pi}{3}\right)^{2/3}\frac{k_{B}^{2}m^{*}n^{1/3}}{\hbar^{2}}
\label{eqn16:gf}
\end{equation}
where k$_{B}$ = 1.38 $\times 10^{-23}$ $JK^{-1}$ is Boltzmann constant, after incorporating the parameter values of $n$, and $\gamma_{n}$, we obtained $m^{*}$ = 5.4(1)$m_{e}$.
The carrier density $n$ is dependent on effective mass $m^{*}$ and Fermi velocity $v_{F}$ by the expression 
\begin{equation}
n = \frac{1}{3\pi^{2}}\left(\frac{m^{*}v_{\mathrm{f}}}{\hbar}\right)^{3}
\label{eqn17:n}
\end{equation}
where $\hbar$ is the Planck's constant, and by employing  $n$ = 4.7$\times 10^{28} m^{-3}$ and $m^{*}$ = 5.4(1) $m_{e}$, it yields $v_{\mathrm{F}}$ = 2.35(6) $\times 10^{5}$  ms$^{-1}$. The mean free path $l$ is related to $m^{*}$, $\rho_{0}$, and $v_{F}$ as 
\begin{equation}
\textit{l} = \frac{3\pi^{2}{\hbar}^{3}}{e^{2}\rho_{0}m^{*2}v_{\mathrm{F}}^{2}}
\label{eqn18:le}
\end{equation}
using the previously calculated parameter values, $m^{*}$ = 5.4(1) $m_{e}$, $v_{\mathrm{F}}$ = 2.35(6) $\times 10^{5} m s^{-1}$, and the $\rho_{0}$ = 71 $\mu\Omega$ cm , we get the mean free path $l$ = 14(1) $nm$. BCS coherence length $\xi_{0}$ can be estimated using the Fermi velocity $v_{F}$ and the transition temperature T$_{C}$ via relation
\begin{equation}
\xi_{0} = \frac{0.18{\hbar}{v_{F}}}{k_{B}T_{C}}
\label{eqn19:f}
\end{equation}
 using $T_{C}$ = 4.0(1) $K$ (from magnetization) and the evaluated $v_{F}$ value gives us $\xi_{0}$ = 84(4) $nm$. We then calculated the ratio of coherence length to the mean free path $\xi(0)$/$l$ = 60(7) >> 1, which indicates the dirty limit superconductivity in HEA sample. The other evaluated parameter are listed in Table \ref{Tab:table1}.\\
 To distinguish MWVRB HEA as conventional or unconventional, we have used Uemura plot classification \cite{U1,U2,U3}. If the ratio of $T_{C}/T_{F}$ for any superconducting material fall in the region 0.01 $\leq$ T$_{c}$/T$_F$ $\leq$ 0.1, then it is considered as the unconventional superconductor, like the organic, heavy fermions and cuprates superconductors. Superconductors falling in the other region with T$_{C}$/T$_F$ $\geq$ 0.1 are classified as the conventional superconductors. To estimate the Fermi temperature value, we have used the expression  $k_{B}T_{F}$ = ${\frac{{\hbar}^{2}}{2m^{*}}(3{\pi}^{2}n)^{2/3}}$ \cite{Fermi Temp}
where $k_{B}$ is the Boltzmann constant, $m^{*}$ = 5.4(1) $m_{e}$, and $n$ = 4.7(2) $\times {10}^{28} m^{-3}$ which yields T$_{F}$ = 9800(357) $K$. The ratio of T$_{C}$/T$_{F}$ is found out to be 0.0004 which places MWVRB HEA far away from the unconventional superconductors \cite {Mayoh,N0.18R0.82_U}. 

\begin{table}[h!]
\caption{Superconducting and normal state parameters of MoReB\cite{MoReB}, WReB \cite{MoReB} and Mo$_{0.11}$W$_{0.11}$V$_{0.11}$Re$_{0.34}$B$_{0.33}$(MWVRB)}
\label{Tab:table1}
\begin{center}
\begingroup
\setlength{\tabcolsep}{4pt}
\begin{tabular}[b]{lcccr}\hline
PARAMETERS& UNITS& MoReB& WReB& MWVRB\\
\hline
\\[0.5ex]                                  
$T_{C}$& $K$& 5.25& 5.30& 4.0(1)\\
$VEC$& & 5.28& 5.28& 5.24\\
$H_{C1}(0)$& $mT$& 13.9& 13.3& 16.8(1)\\                       
H$_{C2}^{mag}$(0)& $T$& 2.1& 2.0& 1.90(1)\\
$H_{C2}^{P}(0)$& $T$& -& -& 7.3(2)\\
$\xi_{GL}$& \text{$nm$}& 12.5& 12.8& 13.1(1)\\
$\lambda_{GL}$& \text{$nm$}& -& -& 160(1)\\
$k_{GL}$& & 15.6& 15.7& 12.2(1)\\
$\Delta(0)/k_{B}T_{C}$& & -& -& 1.81(8)\\
$m^{*}/m_{e}$& & -& -& 5.4(1)\\ 
$\xi_{0}/l_{e}$&   & -& -& 60(7)\\
$v_{F}$& $10^{5}$ $m s^{-1}$& -& -& 2.35(6)\\
$n_s$& 10$^{28}$m$^{-3}$& -& -& 4.7(2)\\
$m^{*}/m_{e}$&  & -& -& 5.4(1)\\
$T_{F}$& $K$    & -& -& 9800(357)\\
$T_{C}/T_{F}$&   & -& -& 0.0004\\
\\[0.5ex]
\hline\hline
\end{tabular}
\par\medskip\footnotesize
\endgroup
\end{center}
\end{table}
 
\section{Conclusion}
In summary, we have synthesized a new superconducting high entropy alloy Mo$_{0.11}$W$_{0.11}$V$_{0.11}$Re$_{0.34}$B$_{0.33}$ and characterized its normal and superconducting state using magnetization, resistivity, and specific heat measurements. It is crystallized in a CuAl$_{2}$ type tetragonal crystal lattice with (I4/$mcm$) space group. The magnetization and specific heat capacity measurements confirm the bulk superconductivity at 4.0 $K$. The electronic heat capacity in the superconducting state is explained well by the conventional BCS model with $\Delta(0)/k_{B}T_{C}$ = 1.81(7) superconducting gap value. The other normal and superconducting parameters of MWVRB are listed in Table \ref{Tab:table1} along with two ternary alloys, MoReB and WReB (crystallized in similar CuAl$_{2}$ type structure). Despite the considerable disorder, MWVRB can be classified as a weakly electron-phonon coupled superconductor and share similarity with ternary CuAl$_{2}$ superconducting family. However, further microscopic measurements such as muon spin rotation/relaxation and theoretical band structure calculations are required to understand the pairing mechanism better. In addition, further superconducting studies on new HEA can help establish a connection regarding the nature of superconducting ground state between ordered and disordered systems.    
\section{Acknowledgments}
Kapil Motla acknowledges the CSIR  funding agency (Award no; 09/1020(0123)/2017-EMR-I), Council of Scientific and Industrial Research (CSIR) Government of India for providing SRF fellowship. R.~P.~S.\ acknowledge Science and Engineering Research Board, Government of India for the Core Research Grant CRG/2019/001028.

\end{document}